\documentclass[12pt, A4]{article}

\usepackage{amssymb, amsfonts, amsmath}

\renewcommand{\cite}[2]{[{#1\if@tempswa , #2\fi}]}

\begin{document}

\title{About some Information and Logical Properties \\ of Semiconductor Multistable Plazma}         
\author{{\small H.~S.~Karayan, Sh.~J.~Martirosyan }\\
{\footnotesize \it Department  of  Physics of  Yerevan  State University }\\
{\footnotesize \it A. Manoogian st.1, Yerevan, Republic of Armenia, 375025}}     
\date{}         
\maketitle

\begin{abstract}
Some information and logical properties of multistable semi-conductor plazma are discussed and the concept of physical logic system $L_p$ on their basis is entered. $L_p$ contains exclusively values of some quantitative characteristic physical size of plazma and transformation of a set of such sizes by means of plazma-field interaction.

The opportunity nonmathematical (physical) calculation in $L_p$ is considered and as an example by symbols $L_p$ are submitted classical arithmetic-logic actions, i.e. their physical models in system $L_p$ are found.
\end{abstract}

\section{Physics in information science}

Quantum physics played rather funny role in  informational - computing technologies. It raised on a qualitatively new level the sphere of implementation of mathematical calculus on the basis of the Boolean logicians. Achievements in the field of technologies, algorithms and the architectures have provided and continue to provide huge successes. But  the quantum physics itself  by its laws braked the powerful triumph of a classic electronics engineering, limiting a capability of mathematical computers and giving rise principled crisis in information technologies. 

The germs of this crisis were pledged also  in the fundaments of the physical theories and have preceded  the  semiconducting electronics engineering, born  by the quantum mechanics. Thus the basis served not only the  boundedness fundamental constant $0< h_1, k, c <\infty$, but also information and logical bases of the physical theories \cite{1-7}. 

Germs of crisis of became more subjected by the appearance  of integral technology and microelectronics, also born by the quantum physics \cite{8,9}.

The completion of this process became the appearance of nonostructure on the stage. It became evident   that if to keep the operational principles of mathematical calculus on the basis of the Boolean logician, some broad classes of problems will be unsolved also in this case \cite{6, 7, 10-13}. 

 As convincingly enough has pointed R. Feynman \cite{6}, this crisis, arisen because of physical limitations,  necessarily must be overcome by the  physics.

From a finiteness of universal  constants the sequence of the following simple facts follows: for implementation of the operation (conversion, carry, processing, storage etc.)   above one bit of the information of a physical system (PS)  the finite time, volume of space, power consumptions and  negentropies are necessary, which  are the characteristics of PS. Any classic association  (i.e. with the classiced correlate parallelism) $m$ such PS for $n$ acts can process the information no more, than $n^{k(m)}$  of bits, thus the characteristic of collective $k(m)<\infty$. It means, that under such prescription only polynomial restricted algorithms are feasible. 

Besides on the basis of the classic (determined) machine it is possible to model classic probabilities, and the quantum probabilities can only be imitated, as legibly enough argued R. Feynman \cite{6,7}. Even for their imitation it is necessary to know beforehand about properties of quantum probabilities, and for modeling it is necessary, that the machine itself had properties of quantum probabilities, i.e. logic nonclassical, accepting a superposition of condition and its interference.

In connection with it, it is possible to distinguish two main reasons, responsible for the currently arisen crisis in informational computing technologies; at first, the information efficiency of the elementary act of calculus is minimum, in second the classic logical basis of calculus itself is minimum.

Both reasons are partially interdependent, therefore clearly or not clearly, they must appear simultaneously. It is visually visible on a model example $q$-bit, which  is represented in PS with two absorbing states $\vert 0\rangle,\, \vert 1 \rangle$   and their coherent superposition (indeterminated condition):
\begin{equation}
\label{eq1}
\vert\alpha\rangle =\alpha_0\vert 0\rangle +\alpha_1\vert 1\rangle;\quad\quad \alpha_0^*\alpha_0 +\alpha^*_1\alpha_1 =1 .
\end{equation}
Because of existing correlation between ?
Here $\alpha$  condition is arbitrary  therefore possesses a huge quantity of information, and its indeterminated  nature is a lien for applying q-bit PS in modeling of quantum probabilities. 

If to consider the conditions $\vert 0\rangle$, $\vert 1\rangle$  and $\vert \alpha\rangle$   accordingly as expressions "lie", "truth", "indefinitely", we shall receive  three-digit quantum logician \cite{2,14}, besides informal in the sense that $\vert \alpha\rangle$  really is true indefinite, and not just by our definition.

It is necessary to mark, that  thanks to $\alpha$ we manage to  receive quantum - PS from $n$ $q$-bits, which is  capable to convert problems with exponential algorithms to polynomial problems. More composite algorithms can be converted to polynomial, if instead of (\ref{eq1}) the  following PS is used:
\begin{equation}
\label{eq2}
\alpha =\sum\limits^k_1\alpha_i\vert i\rangle;\quad\quad\sum\limits^k_i \alpha_i^*\alpha_i=1 .
\end{equation}
Unlike (\ref{eq1}), in (\ref{eq2}) figure $(k-1)$ indefinite elements, therefore the last describes more consistent logical system. This example demonstrates, that the difficulties in information technologies have arisen not because of the physical laws and PS, but because they are not sufficiently used. The informational and logical properties of the nature are much wider, than  the currently used the one-bit information science and Boolean logician. 

In order to use PS in computing process more effectively, we shall return to the analysis of the essence of activity of the mathematical computing (MCM).

The input information - datas and algorithms, is represented on the basis of algebraic pattern. However, MCM executes only logic operations of logical system (Boolean algebra of logicians), therefore it is necessary at first to pass in algebra the logicians (if necessary can be intermediate  transition in a system with binary coding). Later, all the present logic operations in algorithm are   executed,  then all reconversions are made in  converse  order. The logical reasoning or not obviously are present in any link of calculus processing (for example, for compiling algorithm, or some rule, or, program of calculus). But all of them are executed outside of the machine, in which only the logical operations and information transfer are executed. And the mideast role of physics  consists here in implementation of these latests under the simple scheme; to presence of the information carrier (signal) is confronted  the true expression, to the absence - false, and the logical operations (functions) became conversion of input signals in output in accordance with functional settings of these operations. The nature of mathematical calculus is those. But this is not the only way to realize calculus. It is possible to make calculus of other nature, physical, without logical or further without arithmetic operations, and to receive the same result but in the other conception.Two circumstances support this proposal \cite{2,5,14,17};
\begin{enumerate}
\item  Information  quantitative measure of order (on some parameter) PS - is a physical quantity which is capable:
\begin{itemize}
\item to be latent in  system,
\item  is entered in a system and is injected from a system,
\item  is transferred in space and time both inside the system and outside of it,
\item to be converted in a system,
\item  to interact in a system with other informations and physical quantities.
\end{itemize}
\item In PS (and processes in them)  definite logical resources are hidden (logical expressions, formulas, operations etc.) correspondent to the physical laws and legitimacies operating in the system. The information demonstrates a fraction of negentropy which has been turned into the order on the given set of PS microcondition,  and the expression presents  a measure of rejection from chaos at the given microcondition, or, that is equivalent to a measure of veracity (verity) of transformation of the negentropy in the internal order PS in the given microcondition. On the other hand, the information presents a ranked measure of PS and is the quantitative physical characteristic of set of all microcondition, the logical expression quantitatively describes concrete (discrete) condition, introducing a measure and regularity of transformation of  negentropy into the information at the presence of the given condition.
\end{enumerate}

Thus, there are at least  partial conformity between a physical quantity $\xi_p$, describing the given microcondition, by the expression $\xi_{\ell}$, introduced microcondition, and abstract number $\xi_h$, and correspondently by the physical operations above set $\{\xi_p\}$, logic operations above set $\xi_{\ell}$  and algebraic operations on $\{\xi_{h}\}$.

Just the conformity of algebras of quantitative abstract numbers and the logicians have resulted to MCM. In mechanical machines (for example "Feliks", system of interdependent neuron) the arithmetic operatings are made through algebra of logicians located in our mind, i.e. "Feliks" is not the independent machine. The program of calculus  is also stored outside of the machine.

In classic computers the Boolean algebra of logicians is realized, therefore transition to binary algebra is previously committed. Here outside of the machine is the main part of logician, the remaining is inside it.

In both cases the conformity with physics is not used, in view of which it is possible to make calculus, even partially using latent logician of PS.

Moreover, it is possible to take such set of PS, that latent  logician has  ensured the existence of the logical system, which was  selfcontained and full for the given purpose (let's say, for the  implementation of arithmetic operations). The physical quantities (quantum quantity of  microcondition) and physical operators should figure, basically, in this logical system, therefore we shall call physical and we shall designate through $L_p$. $L_p$ not  necessarily should be algebra  of the logicians, but can contain such (for example, Boolean algebra  logicians). By eligible selection PS it is possible to construct algebraic pattern above $L_p$ and to write algorithms of calculus and the applicable machine will be physical.

For more visual demonstrating of these ideas and their reality, below is presented one such example: the simplest  physical logical system multivalued and determined.

\section{Calculus on the basis of multistable plazma   }

The intuitive extension of the logical fundamentals of calculus is connected with transition in the multivalued logician. In order the internal logician is multivalued, it is necessary the  PS to be multistable.

In the nature there are a lot of PS with multirepeatability on any parameter: optical, electrical, $q$-bits, neurons, quantum (Josephson effect, quantum Hall effect and i.e.), each of which can spawn some different logical systems. From behind such outrage more effectively at first  it is purposeful to select satisfactory logical pattern $L_p$, which will dictate itself PS or functional properties.

Let us have different $\xi_i$ of values of convertible value ordered in ascending order and  
$$h_p = \{0,1, \ldots , E\};\quad\quad  E=p-1.$$
The abstract number   $i\in h_p$  indicates both the  value of a physical property  (quantum number) describing $i$ condition, and correspondent logical expression, and also input information condition.

The set $\langle\xi_{1i}, \xi_{i2}, \ldots , \xi_{kik}\rangle$, we call $k$-local $P$-ical  condition (physical, logical, information and algebraic), and bracket $\langle$, $\rangle$, conjunctive, if there is a period $\triangle t\geqslant\tau_{\mbox{\tiny   char}}$ such, that $\forall I\in h_p$ all $\xi_{n i }$ simultaneous (and separately in a case of logician determined considered here) act. Let's designate through $H_k (p)$ set of every possible such condition.

Here $ \tau_{\mbox{\tiny   char}}$  characteristic time  of transition PS from one condition into another. 
We enter also the concept of  alternatively disjunctive bracket $\left[\begin{array}{c} x_i\\ x_j\\ x_m\end{array}\right]$, that $j$ line has place in fulfillment of $j$ condition, which  is incompatible with any $I$ condition at $I\neq j$.    

Set  $\langle [\, ],\ldots , [\,] \rangle$ $m$ - local such brackets we shall designate through $Sh_m$.

As normal subset $H^0_k$  sets $H_k$ we shall call combination of every possible condition $\langle x_1, \ldots , x_k\rangle\in H_k$ such, for which the condition is satisfied:
\begin{equation}
\label{eq3}
\sum\limits^k_{i=1}x_i\leqslant E .
\end{equation}
The definition (\ref{eq3}) of sets of condition $H_k$ decompose to two classes; the class of normal condition $H^0_k$  and its addition to $H_k$.  

Let's define also (logical) norm of condition from $H_k$.
\begin{equation}
\label{eq4}
\vert\langle x_1,\ldots , x_k\rangle\vert  =\min \left(\Sigma x_i, E\right)=
\left[\begin{array}{c}\Sigma x_i\\ E\end{array}\right]=
\left\{\begin{array}{ll} 
\Sigma x_i, & \mbox{if}\; \langle x_1,\ldots , x_k\rangle\in H^0_k\\
E ; & \mbox{if}\; \langle x_1,\ldots , x_k\rangle\not\in  H^0_k. \end{array}
\right.
\end{equation}
The second equaling in (\ref{eq4}) follows from definition (\ref{eq3}).

The norm of a condition introduces a general measure of the verity of the expressions inclusive in this condition, in which  the information k $log_2P$ is also accumulated.

For the basis for construction of the physical determined logical system $L_p$ we shall set elementary physical   transformations of  condition $H_k$  (i.e. we shall define the physical operators above $H_k$).

A). The operation of inverse $I_p\equiv j$  one-one  is convertible  compares to each member $H_1(h_p)$  conjugate (additional) a member from $H_1(p)$;
\begin{equation}
\label{eq5} Ix\equiv E-x;\quad\quad \forall x\in h_p .
\end{equation}

B). Same operation on a subset $\{0, 1\}$  (i.e. Boolean denying):
$$
I_2x =1-x;\quad\quad \forall x =h_2 .
\eqno(5'')$$
And as multi-seater three  imagery   
\begin{equation}
\label{eq6}
1.\hskip50pt K\langle x_1,x_2, x_3\rangle\to I\left\vert \langle x_1, x_2 , x_3\rangle\right\vert\hskip110pt \end{equation}
\begin{equation}
\label{eq7}
2. \hskip50pt h \langle x_1,x_2, x_3\rangle =
\left[\begin{array}{c}E \\ 0\end{array}\right]=
\left\{\begin{array}{ll} 
E, & \langle x_1,x_2, x_3\rangle\in H^0_k\\
0 , & \langle x_1,x_2, x_3\rangle\not\in  H^0_k . \end{array}
\right.\hskip10pt
\end{equation}
$$3.\hskip50pt S \langle x_1,x_2, x_3\rangle =
\left[\begin{array}{c}1 \\ 0\end{array}\right]=
\left\{\begin{array}{ll} 
E, & \mbox{if } \; \langle x_1,x_2, x_3\rangle\in H^0_k\\
0 , & \mbox{if } \; \langle x_1,x_2, x_3\rangle\not\in  H^0_k . \end{array}
\right. \eqno(7')$$

The correlation (\ref{eq5}) - (\ref{eq7}) can be viewed  as definition of functional properties those elementary PS, (the word "elementary" here means, that given PS is one whole, one member (as the transistor)), which are multistable: have multiinput control, and the transformations make by the appropriate  logician.
The right members of definition (\ref{eq5}) - (\ref{eq7}) have describing nature, for example, $x\to I x$, and $I x=E-x$  simple for presentation of properties of $I$ operator on the arithmetic language. In the system $L_p$ there is no necessity for the expression such as ($E-x$ ).

We enter two more "trivial" operators: zero $0$ and identical $1$, which introduce accordingly absence of communications and identical communications  (i.e. the transformation of physical signal or, that is the same, the numbers, information and expression).

Let's make a system $L_p \left(h_p, \leqslant, \langle, \rangle, \left[\begin{array}{c}-\\ -\end{array}\right], 0, 1, I, I_2, S, h, K  \right)$ and  pay attention that the operators on (\ref{eq5}) - (\ref{eq7}) introduce (basic) PS ( we shall call them polistors of the given type \cite{16, 17}), for which all the symbols  in $L_p$ are either the significance  of their characteristic physical parameters, or their property of physical transformation. For example, $I$ - polistor makes $I$ transformations $\forall x\in h_p$ in other member $Ix\in h_p$, i.e. $I x\equiv I(x)$ it is also possible to esteem as a physical quantity,  which is made not through a difference ($E-x$ ).

It is necessary apart to point out, that here sign $"\leqslant "$ (or $""=""$, $"< "$ etc.) introduces not an algebraic logical correlation between the members, but arranges the physical quantities and its operatings are already enclosed in $h_p$, and the sign itself could be excluded from $L_p$, if only we do not want, to  construct  the logician or algebra of logicians on $L_p$.

In order to demonstrate thessence of information and logical  transformations of PS more visually we shall consider a particular example, perhaps $K$-polistor on the basis of multistable semiconducting EHP \cite{8, 16-19}. As was already said, by these conditions were conditioned the latent information and logician polistor. Let the external information be received by the located in non -equilibrium stationary state polister by a conjunctive set of control signals.

The physical nature of the information (and logicians) means, that, apart from energy, definite quantity of a negentropy is received in a system  (proportionally to the norm of the input condition) in the addition  of there existing negentropy. Now in polistor joint actions of both negentropies place joint actions of both negentropies take  place which are capable to induce in the system the new order, if there is such a capability. In EHP the ordering can be made in relation to the contributions of different sorts of electrons and vacant electron sites, say, by relation of the number of electrons passing through a barrier to number of electrons, passing above a barrier; relation of electrons and vacant electron sites were used in formation of different conditions. The ordering on such parameter in space is localized on a place of receipt of the information. At the same time for implementation of transformation of the information with definite logic (regularity), it is necessary, that the proceeding information is spread on all PS. Such information field can be  created in EHP by effect of plasma-field interplay.

After installation of this field, if the norm of  the input condition (the entered negentropy) surpasses some threshold (or a little from them), at the expense of an external power source and entropy in the polistor a new condition is organized and the transformation of the input condition is realized. It is a kind of physical interplay of the external and internal orders and logical expressions with participation of external sources. For $K$ polistor in $H_k$ some physical conditions, due to a permutation symmetry of a conjunctive bracket, are logically indiscernible (i.e. have the identical norm, therefore, describe the identical expressions) and make equivalence classes. If make of these classes the factor set $H_k/L_p$, there will be $P$ members in it, when in $H_k$ exist $P_k$  conditions. And the property of $K$ polistor is to make logic operation (physical transformation of the norms of  condition), is conditioned by that lawfulness (is enclosed in it), according to which  $H_k/L_p$ is imaged in $Sh_m$. $k$-imagery $H_k$ in $Sh_m$, agrees (\ref{eq6}), is not one, but those is the $k$-imagery $H_k/L_p$ in $Sh_1$. The last becomes logically convertible, remaining physically irreversible. Analyzing all characters $L_p$, we come to a conclusion, that apart from customary values, the information and logician have physical nature. In PS,  particularly in polistors, there are quantitative conformity between its physical, information and logical quantitative characteristics and physical operatings above them, and also $L_p$ is algebraic pattern, (but not by algebra), physical quantities and operators.

How much last is valid for representation on its basis the external information, logical, mathematical and physical processes and phenomena?

This problem can be reformulated. If  an abstract  logician  is presented (lets say, Boolean), on the basis of its algebra of  logicians are representable mathematical models therefore physical models are  presented. At the same time are used definite quantity of logical  resources. The problem arises, it is possible to decide a return problem: to present mathematical and logical models through physical with restricted quantity? 

The example $L_p$ is one, but not the only  positive answer on these problems.

To be convinced in it, arithmetic and logic operations universally we represent  through physical quantities and their transformations, (i. e., through characters $L_p$). Their algorithms in the language $L_p$ look like  arithmetic and logic operations. Their algorithms on the language $L_p$ look like:
\begin{equation}
\label{eq8}
\begin{array}{l}
x+y+\eta=\langle K\langle K\langle h\langle Ix, Iy, I_2\eta\rangle, K\langle Ix, Iy, I_2\eta \rangle\rangle, K\langle x, y, \eta\rangle\rangle, S\langle x, y, \eta \rangle\rangle \\
\; \\
x-y=\langle IK\langle K\langle x, Iy\rangle, K\langle Ix, y\rangle, S\langle x, I_2y \rangle\rangle\rangle
\end{array}
\end{equation}

\begin{equation}
\label{eq9}
\begin{array}{l}
x\vee y= IK\langle K\langle Ix, y\rangle ,y\rangle \hskip232pt\\
\; \\
x\wedge y=K\langle K\langle x, Iy\rangle, Iy\rangle\hskip232pt
\end{array}
\end{equation}
The left parts of the formulas (\ref{eq8}) and (\ref{eq9}) present arithmetic and logic operations accordingly, and the extremely physical quantities and transformations figure in the right parts.

As with the help $h_p$ are representable only the ebi-mos of $\sum\limits^n_1 C_i P^i$, where $C_i\in h_p$ that multiplying (and dividing) it is possible to reduce to repeated ($\sum\limits^1_i C_i$  of time) totings (deductions). In (\ref{eq8}) $\eta$ there is transferred unit, that is convenient for applying in computers. If to put $\eta\equiv 0$, we shall receive algorithm of the sum of two numbers. The definitions (\ref{eq9}) provide distributivity of the physical logicians on the basis of $L_p$. It is possible to define them in another way and to receive another logician on $L_p$, including non distributivity. The formulas (\ref{eq8}) and (\ref{eq9}) give an example of physical models, arithmetic and logic operations, and if the algorithms of calculus to translate into the language $L_p$, so  the calculus will be physical, and the realizing machine will be the physical computer.

The element base for such machines, i.e. all the five  basic polistors is possible to realize on the basis of electron-hole plasma, the mathematical model, each of which is possible to present as a  systems from $P$ (model) of different bistable subsystems \cite{18}.
\begin{equation}
\label{eq10}
j=\hat{i}_k x_k+\hat{i}_k\hat{\delta}_k\sqrt{x_k}+\hat{i}_k\hat{\beta}_{k+1}x_{k+1}-j^y_{k},
\end{equation}
\begin{equation}
\label{eq11}
j\left(1-\beta_{k+2}\right)=\hat{\beta}_{k+1}\hat{i}_kx_k+\varphi_{k+1}\left(v_{k+1}\right)+\hat{i}_{k+1} x_{k+1}.
\end{equation}

Where $x_k=(-1)^k\left(1-e^{(-1)^{k+1}\frac{e v_k}{kT}}\right)$ and by $\hat{j}_k^y$, $\hat{i}_k$, $\hat{\delta}_k$, $\hat{\beta}_k$ are denote densitys of management and saturation currents $j^y$ factors of recombination $\delta_k$ and particles transport $\beta_k$ accordingly, with account Plasma-field interaction effect (PFIE) \cite{18-19}.

The outcomes, obtained in activities, and theorems allow to establish not only existence of many solved problems (\ref{eq10}) - (\ref{eq11}) for all polistors $L_p$ with properties (\ref{eq4}) - (\ref{eq7}), but also to select the solution ensuring a processing compatibility of integral fulfillment of polistors in all five  types.

This purpose for electron-hole plasma (EHP) is reached extremely due to operating EPPB in it.

\section*{Conclusion and discussion }

Consideration in the previous section of the elementary models PS-multistable EHP, allows to conclude, that, apart from other physical quantities, the logical expression is also the quantitative characteristic of a system condition which is capable to participate in different interplays in it and to be converted. Thus in  number with energy, entropy etc. definite quantities of the internal latent information and logicians capable partially to be used in information  and computation process inherent in physical systems. In certain conditions resources latent the logicians can be sufficient for construction selfcontained and full, for the  given purpose, physical logical system, most elementary of which one is $L_p$. Determined $L_p$ can become the basis for realization of physical calculus and physical computers. More effective example of physical calculus, are the quantum calculus, marked in section, and computers, in which one will be used latent indeterminated  of the logician for organization of parallel calculus. The essential difference $L_p$ and quantum computers is, that the physical calculus without mathematical and logic operations in a nonclassical logical system is in case of the former made, and in the second case the mathematical calculus in a classic logical system with usage of quantum algorithms is made, i.e. the logical fundamentals of organization of calculus is changed.

The logical properties PS allow to combine these two ways. For $L_p$ this purpose is possible to reach by several images. Most primitive in this schedule, would be usage of neuronic calculus in the neuronic architecture, since polistor $L_p$ are physical neurons, as against programmatic or hardware neurons.

Other capabilities are connected either to the extension $L_p$ up to the quantum logicians, or with quantum - correlated combination of several $L_p$.

All these cases are grounded on latent  logician and on its physical nature.


\end{document}